# Field-free Josephson diode effect in interacting chiral quantum dot junctions

Debika Debnath [1, *] and Paramita Dutta [1, †]

[1] *Theoretical Physics Division, Physical Research Laboratory, Navrangpura, Ahmedabad-380009, India*

We investigate chiral quantum dot (QD)-based Josephson junction and show the correlation-induced Josephson diode effect (JDE) in it. The presence of electron-electron interaction spontaneously creates an imbalance between up- and down-spin electrons during the non-equilibrium transport making the QD effectively magnetic. The simultaneous presence of the chirality and the interaction eventually results in the field-free JDE in our chiral QD junction. We employ the Keldysh non-equilibrium Green's function technique to study the behavior of the Josephson current (JC) and the rectification coefficient (RC) of our Josephson diode (JD). We show a sign-changing behavior of the RC with the Coulomb correlation and the lead-to-dot coupling strength and find the maximum magnitude of the RC ~ 72% for moderate interaction strength. Our proposed field-free JD based on interacting chiral QD may be a potential switching component in superconductor based devices.

## I. INTRODUCTION

Nonreciprocal current flow in heterojunctions has been studied for decades since the development of semiconductor $p$-$n$ junction [1, 2]. With the advancement of the modern technology, several applications have been made using semiconductor materials that carry nonreciprocal currents, like rectifiers, capacitors, switching devices, solar cells, and many more [3, 4]. The recent advent of superconducting diode effect (SDE) [5] has renewed the interest of researchers because of the energy-efficient solution in terms of the dissipationless one-directional supercurrent with potential of applications in nanoelectronic devices [6–15]. Among various superconducting diodes, Josephson diode (JD)s have drawn great attention due to the possibility of tuning the nonreciprocal Josephson current (JC) by the superconducting phase difference externally [16–27]. There are continuous search for possible mechanisms to establish nonreciprocal current in Josephson junction (JJ) with higher and tunable rectification properties.

The origin of nonreciprocal current in JDs lies in the essence of simultaneous breaking of inversion symmetry ($\mathcal{IS}$) and time-reversal symmetry ($\mathcal{TRS}$). Several approaches have been taken in the literature to incorporate the required symmetry-breaking. In order to break the $\mathcal{IS}$, spin-orbit interaction, noncentrosymmetry, and chirality have been mostly considered in various systems of different dimensions like three-layer Rashba superlattice [5], Rashba superconductor (SC) [28, 29], van der Waals material [30], two-dimensional electron gas [31], helical carbon nanotube [32], topological insulator [15, 22], topological semimetal [33, 34], topological SC [35, 36], noncentrosymmetric SC [7, 37], altermagnet junction [12, 38], chiral quantum dot (QD) [25, 26], Rashba spin-orbit coupled QD [26] etc. However, the mutual direction of the Rashba field and the Zeeman field plays an important role here [15, 39].

In addition to the inversion-asymmetry which results in an imbalance between the number of electrons and holes, breaking $\mathcal{TRS}$ is inevitable to separate the number of the up- and down-spin particles required for the nonreciprocity of the JC [40, 41]. The $\mathcal{TRS}$ can be broken in two conceptual ways. The most evident one is the explicit $\mathcal{TRS}$-breaking in the microscopic Hamiltonian. An external Zeeman field, an intrinsic magnetism, or magnetic atomic impurity [15, 28, 42–44] are useful for that. This concept has been utilized in the literature to implement the magneto-chiral anisotropy in both SDE and Josephson diode effect (JDE). Noteworthy, there are some multiband materials with multicomponent superconducting states advocating $\mathcal{TRS}$-breaking alternatively [41, 45–47]. In contrast to the explicit breaking of $\mathcal{TRS}$ in the microscopic Hamiltonian, $\mathcal{TRS}$ can also be broken by other processes. A recent experiment on InAs/GaSb bilayer has explored the Coulomb interaction-induced spontaneous $\mathcal{TRS}$-breaking by tuning the electron and hole densities [48]. Dissipation and electronic correlation have also been utilized for the same purpose [49]. This indirect way of $\mathcal{TRS}$-breaking via correlation is yet to be extensively implemented for the purpose of JDE [40, 50, 51].

One of the finest systems for studying the role of correlations in JDE is the QD-based Josephson junction (JJ). The last two decades have witnessed extensive studies on transport through interacting QD(s) coupled to SC leads [52–64]. The effect of the interplay between the Zeeman field and pair correlations, and the effect of Andreev bound states on the transport have been investigated [65, 66]. To study the correlation effect on the JC, numerical renormalization group (NRG) [52, 67] and functional renormalization group approaches [52], Hartree-Fock (HF) mean-field method [68, 69], Keldysh functional renormalization group [59], Monte Carlo simulations [54] have been used. Another important aspect of Coulomb interaction in QD system is the Kondo effect [70–72] which causes the singlet to doublet phase transition accompanied by $0 - \pi$ junction [59, 67, 73–76].

QD-based junctions are also not exceptions where external Zeeman field or magnetic impurity has been used to break the $\mathcal{TRS}$ of the non-interacting Hamiltonian [25, 26, 44]. On the contrary, in a very recent article, Kocsis et al. have shown zero-field nonlocally tunable SDE in interacting QDs coupled to SCs in 2024 [50]. In another study, diode effect based on the interference effect using two parallel QDs in the JJ with higher rectification coefficient than the value found in the previous JJ [18]. Very recently, Zalom et al. [77] have established the diode effect in an interacting QD for a multi-terminal JJ using the NRG technique. Motivated by these, we consider a gated single interacting chiral QD sandwiched between two SC leads and show correlation-induced JDE in it. The presence of the chirality induces $\mathcal{IS}$-breaking in the system, while, the Coulomb interaction $U$ renormalizes the QD energy $\varepsilon_d$ to $\varepsilon_d + U_{\text{eff}}$ to tunnel electrons from one lead to another [78] with the symmetric Anderson limit (i.e. $\varepsilon_d = -U/2$). The non-equilibrium transport process encodes a finite difference between the up- and down-spin electrons. This correlation-induced number density difference gets coupled to the superconducting phase which regulates the JC. This spontaneous symmetry-breaking during the non-equilibrium transport process in our interacting chiral QD junction results in nonreciprocity i.e., an unequal forward and reverse current ($I_c^+ \neq I_c^-$) is established in the junction [77]. Thus, we show a field-free JDE induced by the intrinsic chirality and the Coulomb correlation. Noteworthy, zero-field JDE has previously been shown in other systems where interaction played a key role [12, 79].

---

* debika@prl.res.in
† paramita@prl.res.in

We use Keldysh non-equilibrium Green's function method to calculate the currents [80]. In the presence of the Coulomb interaction in the QD, three energy scales become competitive in the system: Coulomb correlation, lead-to-QD coupling strength, and the superconducting gap energy [65]. We use the symmetric Anderson model at $\varepsilon_d = -U/2$ and show that the Coulomb correlation acts as a switching parameter to control the sign of the rectification coefficient (RC). To incorporate the effect of the Coulomb correlation we have used the HF mean-field approximation within the prescribed domain of interaction strengths plausible to deal within the mean-field method [81, 82]. We consider only the limit where HF results corroborate with the reults obtained by NRG approach as established by Karrasch et al. [52] and Yoshioka et al. [69] where the Anderson impurity model has been considered to study the JC using both the HF approximation and NRG methods. They showed qualitative similar behaviors of the JCs until the formation of the Kondo singlet.

We organize the rest of the article as follows. In Sec. II, we present the model of our interacting QD-based JJ and the Keldysh Green's function formalism. Our findings are presented and discussed in Sec. III. Finally, we summarize and conclude in Sec. IV.

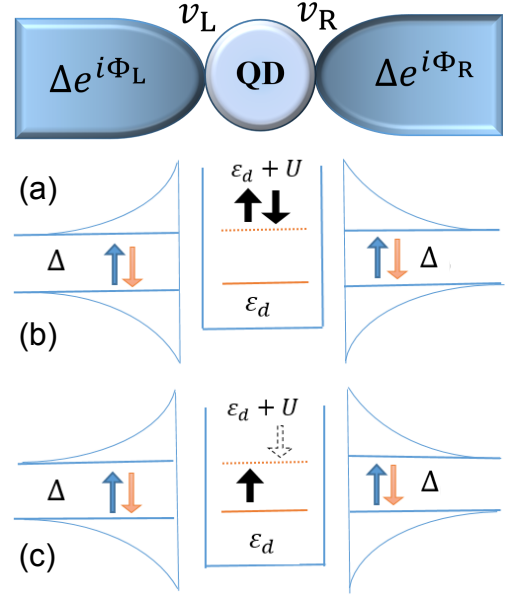

FIG. 1. (a) The schematic diagram of QD-based JJ. QD hosting two electrons at the (b) same site forming a singlet state when $U < v$ and (c) different site forming a doublet state for $U > v$.

## II. MODEL AND FORMALISM

To study the JDE in QD-based junction, we consider the model shown in Fig. 1 where an interacting QD is connected to two SC leads denoted by L and R, respectively. The Hamiltonian for our model is given by

$$\mathcal{H} = \mathcal{H}_{\text{L}} + \mathcal{H}_{\text{dL}} + \mathcal{H}_{\text{dot}} + \mathcal{H}_{\text{dR}} + \mathcal{H}_{\text{R}}, \quad (1)$$

where $\mathcal{H}_{\text{L(R)}}$, $\mathcal{H}_{\text{dot}}$, and $\mathcal{H}_{\text{dL(dR)}}$ represent the superconducting left (right) lead, the QD and the coupling between the QD and the left (right) lead. The terms of Eq. (1) read as

$$\mathcal{H}_{\text{dot}} = \sum_{\sigma}(\varepsilon_{\text{d}} - eV_{\text{g}} + \sigma\alpha\mathcal{I})d_\sigma^\dagger d_\sigma + U d_\uparrow^\dagger d_\uparrow d_\downarrow^\dagger d_\downarrow, \quad (2)$$

$$\mathcal{H}_{\text{L(R)}} = \sum_{\nu \in \{k\text{L}, k'\text{R}\}, \sigma}[\epsilon_\nu a_{\nu\sigma}^\dagger a_{\nu\sigma} + \Delta_\nu a_{\nu\downarrow} a_{-\nu\uparrow} + \Delta_\nu^* a_{-\nu\uparrow}^\dagger a_{\nu\downarrow}^\dagger], \quad (3)$$

$$\mathcal{H}_{\text{dL(dR)}} = \sum_{k\text{L}(k'\text{R}), \sigma}\left[v_{\text{L(R)}} a_{k\text{L}(k'\text{R})\sigma}^\dagger d_\sigma + \text{h.c.}\right]. \quad (4)$$

In Eq. (2), $\varepsilon_{\text{d}}$ denotes the energy level of the QD tunable by an external gate voltage $V_{\text{g}}$ [66]. The current in the JJ induces an intrinsic magnetic field in the system proportional to the current $\mathcal{I}$ that flows through the junction, resulting in a chiral effect in the QD [26, 44] with $\alpha$ being the proportionality constant. The presence of two electrons at the single energy site of the QD costs the Coulomb interaction energy $U$ and the operators $d_\sigma^\dagger(d_\sigma)$ creates (annihilates) an electron in the QD. In Eq. (3), $a_{k\mathrm{L}\sigma}^\dagger (a_{k\mathrm{L}\sigma})$ and $a_{k'\mathrm{R}\sigma}^\dagger (a_{k'\mathrm{R}\sigma})$ are the creation (annihilation) operator for the electron of spin $\sigma$ in the left and right lead with the momentum $k\mathrm{L}$ and $k'\mathrm{R}$, respectively. The onsite energy of electrons in both leads is $\epsilon_\nu$ with $\nu \in \{k\mathrm{L}, k'\mathrm{R}\}$ and the superconducting gap is denoted by $\Delta_\nu = \Delta e^{i\Phi_{\mathrm{L(R)}}}$ where $\Phi_\mathrm{L}$ ($\Phi_\mathrm{R}$) is the superconducting phase for the left (right) lead. The coupling strength between the left (right) superconductor and the QD is described by $v_\mathrm{L}(v_\mathrm{R})$ in Eq. (4), respectively.

Decoupling the superconducting phase $\Phi_{\mathrm{L(R)}}$ and the energy gap $\Delta$, by a canonical transformation [26, 83, 84], we obtain the transformed Hamiltonian as

$$\tilde{\mathcal{H}} = \sum_{\nu \in \{k\mathrm{L},k'\mathrm{R}\},\sigma} \left[\epsilon_\nu a_{\nu\sigma}^\dagger a_{\nu\sigma} + \Delta a_{\nu\downarrow} a_{-\nu\uparrow} + \Delta^* a_{-\nu\uparrow}^\dagger a_{\nu\downarrow}^\dagger\right]$$
$$+ \left[\sum_{k\mathrm{L},\sigma} v_\mathrm{L} e^{\frac{i\Phi_\mathrm{L}}{2}} a_{k\mathrm{L}\sigma}^\dagger d_\sigma + \sum_{k'\mathrm{R},\sigma} v_\mathrm{R} e^{\frac{i\Phi_\mathrm{R}}{2}} a_{k'\mathrm{R}\sigma}^\dagger d_\sigma + \mathrm{h.c.}\right]$$
$$+ \mathcal{H}_{\mathrm{dot}} \quad (5)$$

and calculate the JC in the correlated QD-based junction using the relation given by [85, 86]

$$\mathcal{I} = ie \left\langle \left[\mathcal{N}_\mathrm{L}, \tilde{\mathcal{H}}\right]\right\rangle, \quad (6)$$

where the number of electrons in the left lead is represented by $\mathcal{N}_\mathrm{L} = \sum_{k\mathrm{L},\sigma} a_{k\mathrm{L}\sigma}^\dagger a_{k\mathrm{L}\sigma}$. The explicit form of the JC is obtained as

$$\mathcal{I} = ie \sum_{k\sigma} [v_\mathrm{L} e^{\frac{i\Phi_\mathrm{L}}{2}} \langle a_{k\mathrm{L}\sigma}^\dagger d_\sigma\rangle + (v_\mathrm{L} e^{\frac{i\Phi_\mathrm{L}}{2}})^* \langle a_{k\mathrm{L}\sigma} d_\sigma^\dagger\rangle]. \quad (7)$$

The Coulomb correlation effect is included using the HF mean-field approximation. To ensure the credibility of the approximated method, we scale $U$ and $v$ by $\Delta$ in the considerable limit to avoid the Kondo singlet formation [52, 67]. To find out the JC of Eq. (7), we employ the Keldysh non-equilibrium Green's function formalism as follows. Using the Nambu basis,

$$\psi_\uparrow = \{\epsilon_\uparrow, -\epsilon_\uparrow\}, \; \psi_\downarrow = \{\epsilon_\downarrow, -\epsilon_\downarrow\}, \quad (8)$$

the total Green's function of the QD JJ can be written as,

$$\mathcal{G}^< = \begin{pmatrix} \mathcal{G}_{\mathrm{LL}}^< & \mathcal{G}_{\mathrm{LR}}^< & \mathcal{G}_{\mathrm{L}d}^< \\ \mathcal{G}_{\mathrm{RL}}^< & \mathcal{G}_{\mathrm{RR}}^< & \mathcal{G}_{\mathrm{R}d}^< \\ \mathcal{G}_{d\mathrm{L}}^< & \mathcal{G}_{d\mathrm{R}}^< & \mathcal{G}_{dd}^< \end{pmatrix}, \quad (9)$$

where the off-diagonal terms follow: $\mathcal{G}_{d\mathrm{L(R)}}^< = -(\mathcal{G}_{d\mathrm{L(R)}}^<)^*$. The tunnelling terms of Eq. (7) is contained in the Green's function $\mathcal{G}_{d\mathrm{L}}^<$ which can be represented in the spin $\otimes$ Nambu basis as

$$\mathcal{G}_{d\mathrm{L}}^< = \begin{pmatrix} \mathcal{G}_{d\mathrm{L}\uparrow\uparrow}^< & \mathcal{G}_{d\mathrm{L}\uparrow\downarrow}^< \\ \mathcal{G}_{d\mathrm{L}\downarrow\uparrow}^< & \mathcal{G}_{d\mathrm{L}\downarrow\downarrow}^< \end{pmatrix}. \quad (10)$$

Each component of the Green's function of Eq. (10) can be written in terms of the expectation values of the electronic operators as

$$\mathcal{G}_{d\mathrm{L}\sigma\sigma}^< = i \sum_k \begin{pmatrix} \langle a_{k\mathrm{L}\sigma}^\dagger d_\sigma\rangle & \langle a_{k\mathrm{L}\sigma} d_\sigma\rangle \\ \langle a_{-k\mathrm{L}\sigma}^\dagger d_\sigma^\dagger\rangle & \langle a_{-k\mathrm{L}\sigma} d_\sigma^\dagger\rangle \end{pmatrix}. \quad (11)$$

Therefore, the Keldysh lesser Green's function $\mathcal{G}_{d\mathrm{L}}^<$ is a $4 \times 4$ matrix and the total Green's function $\mathcal{G}^<$ has $12 \times 12$ components. In terms of the Keldysh Green's function, the JC can be written as

$$\mathcal{I} = 2e \sum_{k\sigma} \mathrm{Re}[v_\mathrm{L} e^{\frac{i\Phi_\mathrm{L}}{2}} \mathcal{G}_{d\mathrm{L},\sigma\sigma}^<]. \quad (12)$$

We finally express the JC as a function of the energy given by

$$\mathcal{I} = \frac{e}{\pi} \int \mathrm{Re}[v_\mathrm{L} e^{\frac{i\Phi_\mathrm{L}}{2}} \{\mathcal{G}_{d\mathrm{L},11}^<(E) + \mathcal{G}_{d\mathrm{L},33}^<(E)\}]dE, \quad (13)$$

where $\mathcal{G}_{d\mathrm{L},11}^<$ and $\mathcal{G}_{d\mathrm{L},33}^<$ are the $\uparrow\uparrow$ and $\downarrow\downarrow$ spin electronic components of the Green's function $\mathcal{G}_{d\mathrm{L}}^<$, respectively. Now to calculate the Green's functions explicitly, we use the fluctuation-dissipation theorem [87, 88] in terms of the Keldysh retarded ($\mathcal{G}^r$) and advanced Green's functions ($\mathcal{G}^a$) as: $\mathcal{G}^<(E) = -f(E)(\mathcal{G}^r - \mathcal{G}^a)$, where $f(E)$ is the Fermi distribution function and $\mathcal{G}^a = [\mathcal{G}^r]^\dagger$. Now the Keldysh retarded Green's function can be expressed in terms of the self-energy $\Sigma^r$ and the Green's function $g^r$ of the uncoupled system, using the Dyson equation of motion [89] $G^r = g^r(1 - g^r \Sigma^r)^{-1}$. We refer to Appendix A and Ref. [26] for the details of the analytical calculations. To be noted, the self-energy is a function of the SC phase $\Phi_{\mathrm{L(R)}}$ and the coupling coefficient $v_{\mathrm{L(R)}}$ and $g^r$ contains the Coulomb interaction term. Therefore, the correlation between the Coulomb interaction and the SC phase comes into play in the JC of the heterojunction through the Dyson equation of motion.

To study the rectification quality of our interacting QD-based JD, we define the diode RC as [6, 20],

$$\mathcal{R} = \frac{I_c^+ - |I_c^-|}{I_c^+ + |I_c^-|} \times 100\%. \quad (14)$$

We calculate the JC in units of $e\Delta$ by setting the superconducting phases as $\Phi_\mathrm{L} = -\Phi_\mathrm{R} = \phi/2$ ($\Phi_\mathrm{L} - \Phi_\mathrm{R} = \phi$). Throughout the manuscript, we fix the dot energy $\varepsilon_d = -U/2$ and the chirality coefficient $\alpha = 0.2$ unless specified. The effect of the chirality constant $\alpha$ on the JC and RC is discussed in the Appendix B.

## III. RESULTS AND DISCUSSIONS

Based on the theoretical formalism portrayed in the previous section, we now study the effect of the Coulomb correlation on the behavior of the JC and the rectification property of our QD-based JD. The coupling strength between leads ($v_\mathrm{L}$ and $v_\mathrm{R}$) and the QD acts as an important energy scale competing with the superconducting gap energy ($\Delta$) and the electronic correlation ($U$). We present our results in two sections according to the symmetry between the coupling strengths.

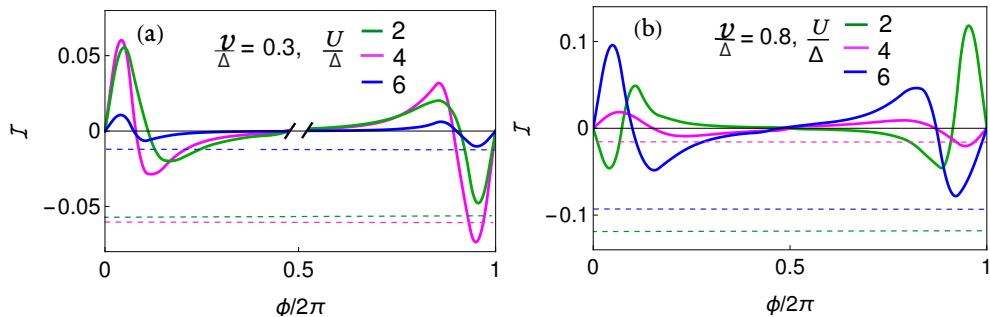

FIG. 2. Josephson current ($\mathcal{I}$) in units of $e\Delta$ as a function of superconducting phase difference ($\phi$) considering (a) $v/\Delta = 0.3$ and (b) $v/\Delta = 0.8$ for different Coulomb interaction strengths ($U$).

### A. Symmetric coupling

In the present subsection, we study the behavior of the JC in terms of the current-phase relation (CPR) and the diode effect for the symmetric coupling strength considering $v_L = v_R = v$.

#### 1. Current-phase relation (CPR)

We start by referring to Fig. 2, where we plot the JC as a function of the superconducting phase difference $\phi$ for two different lead-to-QD coupling coefficients, weak and moderate by setting $v/\Delta = 0.3$ and $v/\Delta = 0.8$ in Fig. 2(a) and (b), respectively, considering various Coulomb interactions. The competition among the lead-to-QD coupling strength, the superconducting gap, and the electronic correlation set constraints on the flow of Cooper pairs through the QD. The correlation resists the formation of Cooper pairs inside the QD, whereas, with the increase in the coupling strength $v$ (comparing Fig. 2(a) and (b)), the JC increases due to the ease of Cooper pairs to tunnel through the junction. The appearance of the $0-\pi$ phase transition in the Fig. 2 for various Coulomb correlations have already been studied in the literature previously [59, 67, 73–76].

In the presence of Coulomb correlation, the discrete energy level in the QD is modified by an additional virtual energy level that appears as a charging energy state $\varepsilon_d + U$ inside the QD. This additional repulsive energy $U$ is the cost that the second electron has to pay in the presence of an electron while transferring through the QD. In the regime when the correlation dominates over the coupling strength ($U/v > 1$), the Coulomb interaction favors each electron to be in the doublet occupancy with the total spin in each level as half. On the other hand, when the effective lead-to-QD coupling becomes stronger than the Coulomb correlation ($U/v < 1$), it allows the QD to have two electrons as a singlet state, rendering total spin zero inside the dot. Depending on the QD energy, the singlet to doublet transition can take place. Thus, the QD energy level also plays an important role.

To enhance the effectiveness of the HF method in the present work, we consider the Anderson particle-hole symmetric model by fixing the QD energy level to $\varepsilon_d = -U/2$. Experimentally, $\varepsilon_d$ can be tuned by an external gate potential [90]. In most semiconductor-based QDs, the electron-hole symmetry is broken due to doping during the fabrication. However, in some materials, like graphene and carbon nanotube QDs, the electron-hole symmetry can be protected [91]. Now, the retro-reflection of the electron as a hole, known as the Andreev reflection, is a key process of the Cooper pair tunnelling through the junction. So, to achieve the higher Cooper pair tunnelling, the Andreev reflection has to be increased, which is possible if the symmetry between the electron and the hole density increases, resulting in a large Andreev reflection. Thus, we consider the symmetric Anderson model, though the effective results will not change qualitatively for other values of $\varepsilon_d$. We also consider the range of electronic correlation $U$ as moderate maintaining $U/v \gg 1$ to ensure that the system remains consistently below the Kondo temperature $T_K$ [92, 93] and the superconducting energy dominates i.e., $T_K/\Delta \ll 1$, where $T_K$ is the characteristic temperature for the Kondo effect calculated in terms of the Coulomb correlation and the coupling strength [67]. We skip the Kondo mechanism in further detail since it is already well-explored in the literature [65, 67, 73] and focus on the diode effect here. We provide a detailed discussion in Appendix C for an estimation of the $T_K$ within the considered energy regimes $U/v \gg 1$ and $T_K/\Delta \ll 1$.

Interestingly, in Fig. 2, we observe that the maximum amplitude of the forward current is not the same as the maximum of the reverse current, i.e., $I_c^+ \neq I_c^-$. We also find that non-sinusoidal behavior appears in the JC whereas, in a regular JJ, the current follows $I = I_c \text{Sin}(\phi)$ relation. Note that, the deviation from the sinusoidal behavior of the current is analyzed in the literature to explore the higher harmonic interferometers [18, 21].

The appearance of the nonreciprocity in the current profile can be explained as follows. The Coulomb correlation in the symmetric Anderson model renormalizes the QD energy and breaks the symmetry of the heterojunction with respect to the Fermi levels in the leads. There is no extrinsic $\mathcal{TRS}$-breaking in the Hamiltonian of our JJ. However, there appears an asymmetry in the number of up- and down-spin electrons ($\langle n_{d\uparrow}\rangle \neq \langle n_{d\downarrow}\rangle$) in the presence of Coulomb interaction making the QD effectively magnetic [69, 94]. This asymmetry affects the JC flow since the number densities are coupled to the superconducting phase factor by Eq. (D1). As a result, phase reversal does not guarantee the same JC in the opposite direction since the $\mathcal{IS}$ is also broken in our chiral QD junction. The asymmetry in the number density enters into our system through the HF approximation of the Coulomb correlation to the Keldysh Green's function following Eq. (D1) (also see Eq. (A8) of Appendix A) which has been already utilized in the literature for studying QD junction. We refer to Appendix D for an estimation of the imbalance between the up- and down-spin electron number densities. Therefore, our JJ shows the diode effect in the presence of interactions that intrinsically separate the up- and down-spin electrons [40, 49] even without breaking the $\mathcal{TRS}$ of the Hamiltonian explicitly. From Fig. 2, it is also clear that the diode effect depends on the interplay of the correlation strength and the coupling constants. A detail investigation of the quality of the diode effect is indeed required from the fundamental as well as the application perspectives. Before we leave the present subsection, one remark is in order. Our present result in the absence of chirality corroborates with the results presented in the supplementary of Ref. [18] in the same limit.

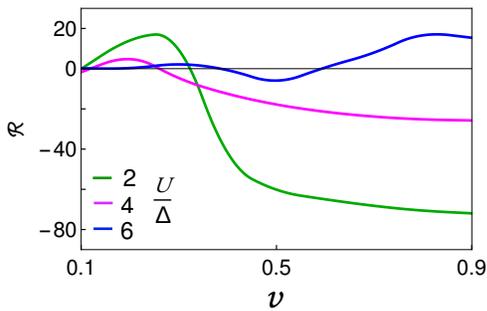

FIG. 3. Rectification coefficient ($\mathcal{R}$%) vs. lead-to-QD coupling strength (in units of $\Delta$) for various correlation strengths ($U$).

#### 2. Diode rectification property

With the discussion of the nonreciprocal behavior of the current in Fig. 2, we now investigate the rectification property of our QD-based JD. Using the definition of Eq. (14) in terms of the forward and reverse current, we plot the RC in Fig. 3 with respect to the lead-to-QD coupling strength $v$ for various Coulomb interaction strengths. Similar to the JC, we ensure to consider various coupling strengths such that $U/v \gg 1$ [52, 69] and $T_K \ll \Delta$ simultaneously for the entire study of the behavior of the RC. In Fig. 3, we see that the behavior of the RC is oscillatory with sign-chaning behavior when the coupling constant increases for a particular $U$. The negative RC indicates a higher reverse current than the forward. It oscillates sign from the positive to negative and vice versa when we change the coupling strength from weak ($v/\Delta \approx 0.1$) to a moderate value ($v/\Delta \approx 0.9$). The sign reversal phenomenon of the RC is obtained as a result of the competition of the three energy scales. Note that, the RC becomes oscillatory for lower values of $v$ mostly. Now, not only the sign, the $U/v$ ratio controls the magnitude of the RC too. The overall behavior of the RC changes with the increase in the correlation strength. We achieve maximum RC as $\mathcal{R} \sim 72\%$ for the correlation $U/\Delta = 2$, which drops down to $\mathcal{R} \sim 20\%$ when the correlation is strong. Note that, in the reality, the Coulomb interaction is constant for a specific material. The coupling strength depends on the fabrication as well as on the SC lead materials. So, in order to utilize this sign-changing behavior and higher RC for practical applications, a combination of JDs based on different materials might be useful. Note that, the amount of the RC will change with the change in the energy level of the QD which can be tuned by the external gate potential as discussed in Refs. [26, 66] in detail.

On the whole, our main observations regarding the characteristic properties of our QD-based JD are twofold: (i) a field-free JDE is achievable for an weak-link (single QD) channel of the JJ in the presence of the chirality and Coulomb interaciton without any explicit $\mathcal{TRS}$-breaking of the Hamiltonian [40, 49] and (ii) the RC (both magnitude and sign) is controllable by the competition between the Coulomb interaction and the lead-to-QD coupling strength.

### B. Asymmetric coupling

With the understanding of the behavior of our QD-based JD for the symmetric coupling strength, we now proceed to explore the same for the asymmetric coupling constant ($v_L \neq v_R$) which is likely to happen in the reality when the junctions are not perfectly hybridized or two SC leads connecting the QD are different. Here also, we consider the parameter regime for the symmetric Anderson impurity model with $\varepsilon_d = -U/2$ where the HF approximation gives comparable results with other established technique like NRG. The asymmetric lead-to-QD coupling indulges a competition with the superconducting gap energy, which eventually helps the screening of the local moments and tries to keep the ground state of the system as a singlet as previously reported in the literature [74]. We consider $v_L$ and $v_R$ such that the system does not host any Kondo singlet and we ensure that by calculating the Kondo temperature $T_K$ (see Appendix C). Thus, we focus on the effect of the asymmetric hybridization on the JC and rectification property.

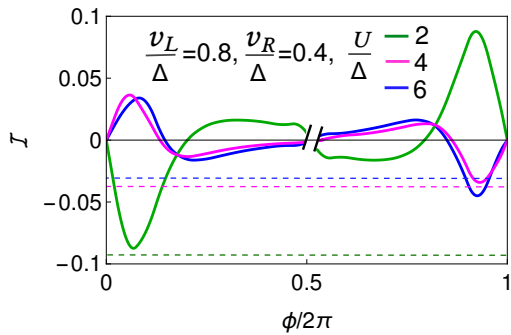
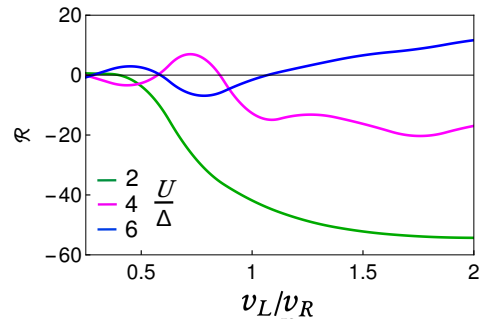

FIG. 4. Josephson current ($\mathcal{I}$) in units of $e\Delta$ as a function of superconducting phase difference ($\phi$) for asymmetric QD-to-lead couplings and various Coulomb interaction strengths ($U$).

FIG. 5. Rectification coefficient ($\mathcal{R}\%$) vs. the ratio of the two lead-to-QD coupling strengths ($v_\text{L}/v_\text{R}$) for various correlation strengths ($U$).

#### 1. Current-phase relation (CPR)

In Fig. 4, we show the effect of asymmetric coupling on the JC for various Coulomb correlation strengths. With the increase in the Coulomb correlation, the JC is reduced due to the Coulomb repulsion. We find that the asymmetry in the coupling coefficients in the two leads enhances the asymmetry in the forward and reverse currents ($I_c^- \neq I_c^+$). We observe that the JC changes sign with the change in the Coulomb correlation strength from $U/\Delta$. This sign change in the JC is an artefact of the $0-\pi$ phase transition in the JJ, which is well-explored in the literature [50, 54, 55, 95, 96]. For moderate Coulomb interaction strength $U/\Delta = 4$, the forward current is higher than the reverse current ($I_c^+ > I_c^-$), whereas, for weak and comparatively larger $U$ values, the reverse current dominates the forward current, resulting in an efficient diode effect.

#### 2. Diode rectification property

In order to investigate the effect of the asymmetry of the coupling constants on the rectification, we now present the behavior of the RC as a function of the ratio of the lead-to-QD coupling $v_\text{L}/v_\text{R}$ for various correlation strength in Fig. 5. We find that the RC changes sign as the asymmetry $v_\text{L}/v_\text{R}$ increases. The change in the sign of the RC from the negative to the positive implies that the reverse current is higher than the forward current which is switched to the opposite i.e., the forward current becomes higher than the reverse one as mentioned in the previous section. Note that, in the previous section, the sign-chaning behavior is shown as a function of the correlation strength, which is true for the asymmetric coupling too. We avoid showing similar results for the asymmetric coupling case. Additionally, we find the sign-chaning behavior here with the asymmetry of the lead-to-QD coupling as well. Not only the sign, the magnitude of the RC also varies with the asymmetry depending on the interplay of the strength of the asymmetry and the correlation. We find higher RC ($\mathcal{R} \sim 60\%$) with strong asymmetry in the coupling strength for weak correlation, whereas, the RC drops to $\mathcal{R} \sim 12\%$ when both the correlation and the asymmetry are strong. Thus, an optimum condition for the Coulomb correlation and the junction coupling strength is required to maximize the RC.

### IV. SUMMARY AND CONCLUSION

In the present work, we have studied a chiral QD JJ in the presence of electron-electron interaction inside the dot. The number of up-spin and down-spin electrons inside the QD also differs due to the effective electronic correlation in the non-equilibrium transport. We solved the Coulomb correlation problem at the mean-field limit maintaining $T_K \ll \Delta$ and calculated the JC using the Keldysh non-equilibrium Green's function technique. We have found that the forward and reverse current in the JJ are significantly different from each other for both the symmetric and asymmetric coupling strengths of the two leads. The rectification property of our JD strongly depends on the competitive effects of the electronic correlation and the lead-to-QD coupling strengths. The sign-changing behavior of the RC depends on the Coulomb correlation strength in the QD and the lead-to-QD coupling coefficients. In all the previous works on the QD-based JDs, either an external magnetic field or magnetic impurity has been considered [25, 44]. The recent experimental result on the JDE in a magnetic atom [43] and its consecutive theoretical model with a magnetic impurity in a QD [44] support our result of the JDE. The uniqueness of our work lies into showing the appearance of the diode effect in the absence of any external field or magnetic impurity i.e., without explicitly breaking the $\mathcal{TRS}$ in the microscopic Hamiltonian of the chiral QD. We have computed all our numerical results at zero temperature. Finite temperatures (necessarily less than the critical temperature) may cause thermal fluctuations, resulting in a relatively lower diode effect at high temperatures. At high tempereatures, usage of high $T_c$ superconductor may be helpful [97].

In the context of experimental realization, there has been a significant boost in the experimental research of JJ with QD following the upsurge of theoretical study during the last two decades, [73, 78, 98–100]. For our proposed model, an asymmetric junction can be realized either by considering two different superconducting materials and/or different couplings. The chiral QD can be formed by using strain or distorted structure [101–103]. Recent work on the synthesization of chiral QD based on the chiroptical spectroscopy has proposed a new way for chiral QD characterization [104]. Maintenance of an optimum correlation strength can be controlled by fabricating the QD with the optimum radius, so that the junction does not reach a Coulomb blockade regime [105, 106]. Considering a charging energy $E_c$ of the QD could resemble a capacitor circuit as explained in the Ref. [107]. Tunability of the QD energy is possible by using an external gate voltage [26, 90, 108]. Josephson diode experiment on a single magnetic (using Pb, Cr and Mn) atom shows the possibility of realization of the Josephson current in the QD structure [43]. The work on the transmon circuit by a gate-defined QD of an InAs-Al nanowire-based JJ depicts the model of a single-impurity Anderson model with superconducting leads [76].

In reality, the Coulomb correlation depends on the size of the QD as $U = \frac{e^2}{4\pi\epsilon_0 \epsilon r}$, where $e$ and $\epsilon_0$, $\epsilon$, $r$ are the electronic charge, vacuum permittivity, effective dielectric constant of the material or medium, and the diameter of the QD, respectively. For a specific material, $\epsilon$ is constant but the QDs can be fabricated with different sizes so that the electronic interaction in the QD can be controlled. For example, in semiconductor QDs like InAs and GaAs QDs of sizes ~ 10 nm, the Coulomb correlations $U$ ~ 19 meV and ~ 112 meV, respectively. For a carbon nanotube, it becomes $U$ ~ 72 meV keeping the same size. Thus, the correlation is defined by the choice of the material of the QD and its size. Controlling the QD size provides the freedom to fabricate the heterojunction with its highest efficiency [109]. For SC leads, Nb or Al with the transition temperature $T_c$ ~ 10 K, the gap $\Delta_0$ ~ 1.41, and 0.18 meV, respectively, can be considered. There exists a wide range of possibilities [110, 111] for choosing SC materials in the JJ.

A few comments on the comparison of our results with the existing literature are in order. The main advantage of our work lies into the fact that we do not need to apply any external field to achieve the diode effect, whereas, in the literature most of the works are based on either the Zeeman field or magnetic impurity. Also, most of the JDE results are shown in the noninteracting model. So, our model of JDE is much more realistic in that sense. The recent work on the Josephson junction with a magnetic impurity in a QD structure shows the diode rectification coefficient is tunable with the magnetic moment of the impurity and there is an optimum strength of the moment that gives the highest rectification [44]. This result that a moderate Coulomb correlation gives the highest RC and beyond that limit (here $U = 2\Delta$ in our system) it reduces, corroborates our findings. The advantage of our model over this work is that we do not need to add any magnetic impurity here. Also, we have achieved higher RC (~ 72%) compared to the maximum RC shown in Ref. [44] (~ 58%). To compare with the other existing results in QD based JDs, in Ref. [18], Souto *et al.* showed a quantum interferometer using two correlated parallel QDs and found RC up to 45%. While, Zalom *et al.* showed the diode effect in a multiterminal JJ based on interacting QD with a finite RC in the JC [77]. Although, they did not show RC explicitly, but the magnitude (RC ~ 60%) can be extracted from their results of JC. In another work, Pillet *et al.* showed diode effect in an Andreev molecule in the presence of Coulomb interaction that gives rise to RC~ 25%[112]. In non-interacting QD JJ in the presence of external magnetic field, a maximum of RC~ 12% has been shown [25]. Thus, to the best of our knowledge, we have achieved the highest JDE so far among all QD based JJs. The effect of the competition between the Coulomb interaction, the junction coupling strength and the SC gap energy simultaneously play a significant role in the JDE. This makes our QD-based JD model very efficient for studying the spin-qubits and quantum interference device (SQUIDS) [21, 113–116], quantum interferometers [18, 117], single-electron transistors [106, 118, 119] and many more.

## ACKNOWLEDGMENTS

We acknowledge the Department of Space (DoS), Government of India for all support at Physical Research Laboratory. D. D. thanks Kuntal Bhattacharyya for bringing some important references into notice and Diptiman Sen for helpful discussions. P. D. acknowledges Mohit Randeria, Swati Chaudhary, and Debmalya Chakraborty for helpful discussion and the Anusandhan National Research Foundation (ANRF) (erstwhile Science and Enghineering Research board (SERB)), Department of Science and Technology (DST), Government of India, for the financial support through the Start-up Research Grant (File no. SRG/2022/001121).

## Appendix A: Keldysh Green's function and Hartree-Fock method

In this section, we describe the HF solution of the Coulomb correlation and some important steps of the Keldysh Green's function method to calculate the JC through the correlated JJ.

In the Anderson impurity model, the effect of the Coulomb correlation can be calculated using the HF approximation when the spins of all the impurities in the system have a significant localized effect and the excitations in the system could be considered in the mean-field limit. In the HF regime, the model Hamiltonian of Eq. (2) can be written as,

$$\begin{aligned}\mathcal{H}_{\text{dot}} &= \sum_\sigma \varepsilon_{\text{d}} d_\sigma^\dagger d_\sigma + U(\langle n_{d\downarrow}\rangle n_{d\uparrow} + \langle n_{d\uparrow}\rangle n_{d\downarrow} - \langle n_{d\downarrow}\rangle\langle n_{d\uparrow}\rangle) \\ &= \sum_\sigma (\varepsilon_{\text{d}} + U\langle n_{d\bar\sigma}\rangle) d_\sigma^\dagger d_\sigma - U\langle n_{d\downarrow}\rangle\langle n_{d\uparrow}\rangle. \quad \text{(A1)}\end{aligned}$$

The electron number operators $\langle n_{d\sigma}\rangle$ are calculated self-consistently. The accuracy of the JC calculated using the HF method depends on the choices of several energy parameters in the system. We consider the Coulomb correlation and the lead-to QD coupling parameters such that $U/v \gg 1$ maintaining $T_K \ll \Delta$ throughout our work.

Considering the HF approximated Hamiltonian, we calculate the JC using the Keldysh non-equilibrium Green's function technique. In order to solve the Hamiltonian, we use the Dyson equation of motion: $G^r = g^r + g^r \Sigma^r G^r$ to calculate the self-energy of the system in terms of the Keldysh retarded Green's function[89]. We find the retarded self-energy using the basis of Eq. (8) of the main text as

$$\Sigma^r = \begin{pmatrix} 0 & 0 & V_L \\ 0 & 0 & V_R \\ V_L^* & V_R^* & 0 \end{pmatrix}, \quad \text{(A2)}$$

where $V_{L(R)}$ represents the normalized coupling strength between the left (right) superconductor. This can be written as

$$V_{L(R)} = \begin{pmatrix} \mathcal{V}_{L(R)} & 0 & 0 & 0 \\ 0 & -\mathcal{V}_{L(R)}^* & 0 & 0 \\ 0 & 0 & \mathcal{V}_{L(R)} & 0 \\ 0 & 0 & 0 & -\mathcal{V}_{L(R)}^* \end{pmatrix}, \quad \text{(A3)}$$

where $\mathcal{V}_{L(R)} = v_{L(R)} e^{i\frac{\Phi_{L(R)}}{2}}$. The Keldysh retarded Green's function for the uncoupled leads and the QD read as

$$g^r = \begin{pmatrix} g_{LL}^r & 0 & 0 \\ 0 & g_{RR}^r & 0 \\ 0 & 0 & g_{dd}^r \end{pmatrix}. \quad \text{(A4)}$$

The retarded Green's function for each lead of Eq. (A4) can be written as

$$g_{\nu\nu}^r = \begin{pmatrix} \langle\psi_\uparrow|g_{\nu\nu}^r|\psi_\uparrow\rangle & I_2 \\ I_2 & \langle\psi_\downarrow|g_{\nu\nu}^r|\psi_\downarrow\rangle \end{pmatrix}, \quad \text{(A5)}$$

where $\nu \in L/R$, $I_2$ is the $2\times 2$ identity matrix, and $\psi_\sigma$s are given by Eq. (8) of the main text. The other matrix elements are given by $g_{\nu\nu,\epsilon\sigma,\epsilon\sigma}^r = -i\pi\rho\rho_\nu$ and $g_{\nu\nu,\pm\epsilon\sigma,\mp\epsilon\sigma}^r = -i\pi\rho\rho_\nu \sigma\Delta/(\epsilon + i\eta^+)$. We consider the density of states for the QD as unity and the leads as

$$\rho_\nu = \begin{cases} \frac{|\epsilon_\nu|}{\sqrt{\epsilon_\nu^2 - \Delta^2}} & |\epsilon_\nu| > \Delta \\ \frac{-i\epsilon_\nu}{\sqrt{\Delta^2 - \epsilon_\nu^2}} & |\epsilon_\nu| < \Delta. \end{cases} \quad \text{(A6)}$$

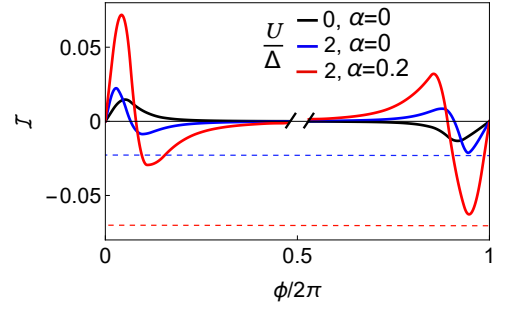

FIG. 6. Josephson current ($\mathcal{I}$) in units of $e\Delta$ as a function of superconducting phase ($\phi$).

The Green's function of the uncoupled QD with the electronic interaction is written as

$$g_{\text{dd}}^r = \begin{pmatrix} g_{\text{e},\uparrow} & 0 & 0 & 0 \\ 0 & g_{\text{h},\uparrow} & 0 & 0 \\ 0 & 0 & g_{\text{e},\downarrow} & 0 \\ 0 & 0 & 0 & g_{\text{h},\downarrow} \end{pmatrix}, \quad \text{(A7)}$$

where each term follows,

$$g_{\text{e(h)},\sigma} = \frac{\tilde\omega_{\text{e(h)},\sigma} - U + U\langle n_{\text{d}\bar\sigma}\rangle}{\tilde\omega_{\text{e(h)},\bar\sigma}(\tilde\omega_{\text{e(h)},\sigma} - U)}, \quad \text{(A8)}$$

with $\tilde\omega_{\text{e},\sigma} = \epsilon_\nu + i\eta - \varepsilon_{\text{d}} + \sigma\alpha\mathcal{I}$ and $\tilde\omega_{\text{h},\sigma} = \epsilon_\nu + i\eta + \varepsilon_{\text{d}} + \sigma\alpha\mathcal{I}$. The contribution by the up- and down- spin electrons to the Green's function is decided by the combined effect of $U$ and $\alpha$. The sign of the difference between the up- and down- spin number density $(\langle n_{d\uparrow}\rangle - \langle n_{d\downarrow}\rangle)$ and the field induced by the chirality will be different for the two spins along the forward and the reverse current direction as follows.

| Direction of Current | ↑ spin | ↓ spin |
|---|---|---|
| Forward | $\delta n + \alpha I$ | $-(\delta n + \alpha I)$ |
| Reverse | $\delta n - \alpha\ I$ | $-(\delta n - \alpha I)$ |

Using the Keldysh retarded Green's function $g^r$ of the uncoupled system and lead-to-QD coupling matrices $(V_L, V_R)$, self-energy of the system $\Sigma^r$ is calculated. Then, the Keldysh lesser Green's function $\mathcal{G}^<$ is calculated numerically using the fluctuation-dissipation theorem.

### Appendix B: Effect of chirality

We now discuss the effect of the chirality in detail. This chirality in the QD can be induced by several ways, like preparing an asymmetric QD structure or by using strain that may cause an asymmetric band spectrum in the system similar to the asymmetric band structure considered in the literature [120]. Chiral QDs have been shown to have a series of applications in the field of quantum technologies and device fabrication [101–104].

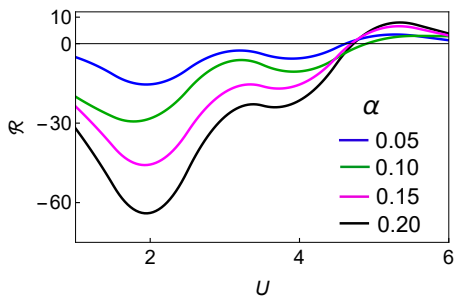

FIG. 7. RC as a function of correlation strength $U$ (in units of $\Delta$).

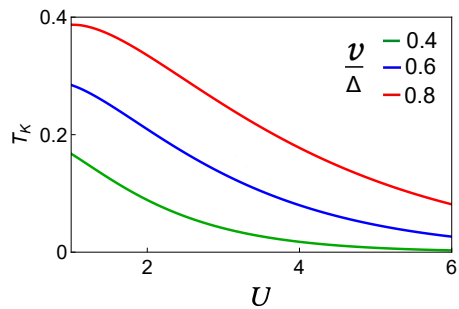

FIG. 8. Kondo temperature ($T_K$) as a function of correlation strength $U$ (in units of $\Delta$).

In our model, the JC flow induces a magnetic field which is proportional to the current ($\sigma \alpha I$ term in the Hamiltonian of Eq. (2)). The forward current will experience the geometry with the opposite helicity compared to the reverse current [25]. The opposite chiral current in the QD will induce the field along the opposite direction. The field realized by the up/down spin electron ($\sigma = \pm 1$) in the system due to forward (reverse) current is $\sigma(\pm \alpha |I|)$. We have solved this self-consistently. To understand the effect of the chirality more clearly, we plot the JC in Fig. 6 for zero and finite chirality constant both in the presence and absence of the correlation. We find that the JC increases with the increase in the Coulomb interaction, but it increases more when the chirality coefficient becomes finite. In the absence of Coulomb correlation and chirality, the forward and the backward currents are exactly similar to each other i.e., $I_c^+ = I_c^-$. When we introduce the chirality in the QD, the finite Coulomb interaction gives rise to a vanishingly small asymmetry in the currents. This result in the absence of chirality corroborates with the results presented in the supplementary of Ref. [18] in the same limit of the correlation. The difference between $I_c^+$ and $I_c^-$ increases with the finite chirality. The forward current is higher than the reverse current in the presence of the chirality and the correlation. We show the results for the lead-to-dot coupling $v/\Delta = 0.6$, but the asymmetry among the currents remains valid qualitatively for other values too. However, whether the forward current will be higher than the reverse or not will depend on the strength of the lead-to-QD coupling strength as shown in the main text. In Fig. 7, we show the RC for different chirality coefficients and find that the RC increases with the chirality. We also observe that the moderate Coulomb correlation is more efficient in increasing the current nonreciprocity.

### Appendix C: Role of Kondo effect

In this section we present an account of the well-known Kondo effect which plays an important role in the transport through QD JJ. In an interacting QD-based JJ, the Kondo effect arises if the strong correlation through the formation of the Kondo singlet dominates the superconductivity. The competition between different energy scales like the superconducting gap energy $\Delta$, Coulomb correlation $U$ and lead-to-dot coupling $v$ play significant roles in determining the state of the many-body system. To use the HF mean-field method for calculating the JC, it is important to prepare the system so that the Cooper pair tunnelling is not affected by the Kondo effect [52, 67, 69]. To ensure it, we measure the Kondo temperature of the system which depends on the $U, v$ and the QD energy level $\epsilon_d$ following the relation [67, 92],

$$T_{\mathrm{K}} = \sqrt{\frac{Uv}{2}} exp[\pi \frac{\varepsilon_d}{2v}(1 + \frac{\varepsilon_d}{U})]. \quad \text{(C1)}$$

and plot it in Fig. 8, we plot it as a function of correlation strength for various lead-to-QD coupling coefficients $v$ maintaining $T_K \ll \Delta$. The decaying nature of the Kondo temperature proves that the system does not enter the Kondo regime for the entire regime of $U$.

### Appendix D: Electron number densities

For an interacting QD, the Coulomb correlation using the HF approximation originates an intrinsic magnetic moment due to the difference in the number of up- and down-spin electrons and this finite magnetic moment orients along the axis dominated by the relative spin polarization [69, 94]. To find out the intrinsic reason for the underlying diode effect in our model JJ, we calculate this change in the number of the up- and down-spin electrons ($\delta n = |\langle n_{d\uparrow}\rangle - \langle n_{d\downarrow}\rangle|$) in the QD. Using Eq. (13), we calculate the spin-dependent electron densities self-consistently following the relation

$$\langle n_{\mathrm{d}\sigma}\rangle = \int \frac{dE}{2\pi i} \mathcal{G}^<_{\mathrm{dd},\sigma\sigma}(E) \quad \text{(D1)}$$

and plot it in Fig. 9 considering different values of $U$ keeping $v/\Delta = 0.6$. We see that the number densities for the up- and down-spins are not the same. The difference i.e., $\delta n$ varies with $\phi$ and also with $U$. This change in the number of up- and down-spin electrons generates the magnetic moment in the QD which modifies with the

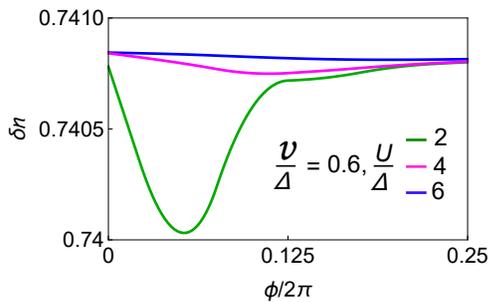

FIG. 9. Difference in up-spin and down-spin electron number densities $\delta n$ as a function of superconducting phase ($\phi$).

Coulomb correlation in a similar way as the Coulomb impurity in the Anderson model [94].